\documentclass[12pt,a4paper]{article}
\usepackage{amssymb,amsmath,amscd,epsfig}
\title{Possible violation of the spin-statistics relation for neutrinos: 
cosmological and astrophysical consequences}
\author{
A.D. Dolgov$^{\rm a,b,c}$, A.Yu. Smirnov$^{\rm c,d}$, 
\\[5mm]
${\rm ^a}$ {\small\it Istituto Nazionale di Fisica Nucleare,
Ferrara 40100, Italy} \\
${\rm ^b}$ {\small\it Institute of Theoretical and Experimental Physics, 
117218, Moscow, Russia}\\
${\rm ^c}$ {\small\it 
The Abdus Salam International Centre for Theoretical Physics, 
I-34100 Trieste, Italy}\\
${\rm ^d}$ {\small\it Institute for Nuclear Research, Russian Academy of
Sciences, Moscow, Russia} 
}
\date{}

\begin{document}

\newcommand{\be}{\begin{eqnarray}}
\newcommand{\ee}{\end{eqnarray}}
\newcommand{\bi}{\bibitem}
\newcommand{\lar}{\leftarrow}
\newcommand{\rar}{\rightarrow}
\newcommand{\lrar}{\leftrightarrow}
\newcommand{\mplq}{m_{Pl}^2}
\newcommand{\mnu}{m_\nu}
\newcommand{\nnu}{n_\nu}
\newcommand{\ngam}{n_\gamma}
\newcommand{\ms}{m_S}
\newcommand{\taus}{\tau_S}
\newcommand{\ns}{n_S} 
\newcommand{\me}{m_e}
\newcommand{\dnnu}{\Delta N_\nu}
\newcommand{\Tbbn}{T_{BBN}}
\newcommand{\nue}{\nu_e} 

\maketitle

\begin{abstract}

We assume that the Pauli exclusion principle is violated for 
neutrinos, and consequently, neutrinos obey the Bose-Einstein 
statistics. Cosmological and astrophysical consequences of 
this assumption are considered. Neutrinos may 
form cosmological Bose condensate which accounts for all (or 
a part of) the dark matter in the  universe. ``Wrong'' statistics 
of neutrinos could modify big bang nucleosynthesis, leading to the 
effective number of neutrino species smaller than three.  
Dynamics of the supernova collapse would be influenced and 
spectra of the supernova neutrinos may change. The presence of 
neutrino condensate  would enhance contributions of the Z-bursts  
to the flux of the UHE cosmic rays and lead to substantial 
refraction effects for neutrinos from remote sources. 
The Pauli principle violation for neutrinos can be tested in 
the two-neutrino double beta decay.

\end{abstract} 

\section{Introduction \label{s-intro}}

What is the next surprise neutrinos will bring us? Pauli  has  
introduced neutrino to resolve paradoxes of the beta decay, in 
particular, an apparent violation of the spin-statistics relation. 
Could the neutrino itself violate this relation? 
Does the particle invented by Pauli respect the Pauli principle? 
Do we have any indication to that? 

The puzzle of cosmological dark matter (DM) remains with us already for more
than a half of century but we still do not know what are the constituents 
of this mysterious substance. The commonly accepted point of view is that 
dark matter is made of new elementary particles governed by the laws
of the old established physics. Here we will explore a different 
possibility: old particles and new physics. Namely we assume that Fermi
statistics for neutrinos is violated and, if so, neutrinos with 
a fraction of eV masses, as observed in the oscillation experiments,
could condense and make all cosmological dark matter, both cold and 
(a little) hot.\\

We  suggest that Pauli exclusion principle is violated for neutrinos 
and therefore neutrinos obey (at least partially) the Bose-Einstein 
statistics.  Of course, a possibility to explain the dark matter is 
not the only consequence  of the spin-statistics 
violation. One can expect some effects of the violation in any environment 
where large densities or fluxes of neutrinos exist. That includes 
the Early Universe in the epoch of the Big Bang nucleosynthesis, 
the cores of collapsing stars, {\it etc.}.  Some consequences could be 
also seen in laboratory experiments (e.g.,  in the double beta decay). 

Possible violation of the exclusion principle was discussed 
in a series of theoretical papers~\cite{ign-kuz} though no satisfactory 
model has been proposed so far. (For a critical review see 
ref.~\cite{lbo-rev}.) Experimental searches of the
Pauli principle violation for electrons~\cite{exp-viol}  
and nucleons~\cite{exp-bar} have given negative results.  
It may happen however that neutrinos 
due to their unique properties are  much more sensitive to the 
violation and it is in the neutrino sector the effects can  be 
seen first. 
Neutrinos may also possess kind of mixed or 
more general statistics than Bose or Fermi ones~\cite{para}.

The assumption of violation of the Pauli exclusion principle 
reveals immediately a number of problems. The spin-statistics 
theorem follows from the canonical quantization to ensure a positive 
definiteness of energy. It is not clear how to overcome this
problem and how serious it is for neutrinos. 
The CPT theorem follows, in particular, from the normal 
relation between spin and statistics, therefore the suggested scenario  
may also violate the 
CPT theorem. Actually, a  possible violation of the latter in neutrino 
physics is under an active study now, see, {\it e.g.}, refs.~\cite{cpt}. 
Last but not least, the spin-statistics violation in the neutrino sector 
is communicated due to the weak interactions to 
charged leptons and other fermions where the bounds are extremely 
strong. It is not clear if effects considered in this paper are 
consistent with these bounds, which depends on particular mechanism 
of the violation.

In what follows we put aside discussion of these problems. 
Instead, taking pure phenomenological approach, 
we concentrate on cosmological and astrophysical consequences 
of the neutrino ``bosonization''    
in an attempt to find interesting observable effects or to restrict 
such a possibility.

\section{Bosonic neutrinos: Context}

The standard electro-weak theory puts the left-handed
neutrinos and electrons into the same doublet and thus one would expect
that neutrinos and electrons obey the same statistics. On the other
hand, as we know,  being the only neutral leptons, the neutrinos can have 
substantially different properties from those of the charged leptons. 
In particular, neutrinos can be the Majorana particles and 
induce the lepton number 
violation. The difference between  the charged leptons and neutrinos 
is related to  breaking of the electro-weak (EW) symmetry.
The lepton number violation (in the context of seesaw mechanism) 
originates from very high scales.

Similarly, the neutrino sector might be a source of  violation
of the spin-statistics relation;  this can also be  
connected to  EW symmetry breaking and originate from  some high mass  
scale of Nature.  One may consider  scenario where  violation of the Pauli 
principle occurs in  a hidden sector of  theory related to 
the Planck scale physics, or strings physics. 
It could be mediated by some singlets of the Standard model - (heavy) neutral 
fermions which mix with neutrinos when the EW symmetry is broken. 
Since only  neutrinos can mix with the singlets, effects of the Pauli 
principle violation would be manifested first in neutrinos and then 
would communicate to other particles. 
Also one can consider a possibility that the messenger of 
the Pauli principle violation is the light sterile neutrino. 
It has a small mixing with the active components, and  this small mixing  
quantifies the degree of violation in the observable sector.  
In this way a small or partial violation of relation between spin and 
statistics  might occur. 

As in the case of lepton number, a violation of the 
spin-statistic relation for  other particles can be suppressed by 
an additional power of a small parameter relevant for the violation 
in the neutrino sector. 
In fact, the high accuracy of the validity of Fermi statistics for 
electrons~\cite{exp-viol}  
could put a strong bound on a possible ``transfer'' of wrong statistics 
from neutrinos  to electrons. 
In this connection one can consider 
a scenario when significant effects of the spin-statistics 
violation develop on the cosmological times or in particular 
environments of the Early Universe.\\ 

A violation of the Pauli principle for neutrinos
should show up in the elementary processes where identical
neutrinos are involved. A realistic process for  this test  
is the two-neutrino 
double beta decay, $A\rar A'+ 2\bar{\nu} + 2e^-$ (or similar with 
antineutrinos and positrons). The  probability
of the decay as well as  the energy spectrum
and angular distribution of electrons should be   
affected.

The difference between neutrino statistics would be strongly pronounced 
if the probabilities  (see, {\it e.g.}  
\cite{boehm}) are proportional 
to  the bi-linear combinations  of the type
$K_m K_n$,  $K_m L_n$, $L_m L_n$, where
\begin{eqnarray}
K_m \equiv [E_m  - E_i + E_{e1} + E_{\nu 1}]^{-1} - [E_m  - E_i + E_{e2} + 
E_{\nu 2}]^{-1},
\nonumber\\
L_m \equiv [E_m  - E_i + E_{e2} + E_{\nu 1}]^{-1} - [E_m  - E_i + E_{e1} + 
E_{\nu2}]^{-1},
\label{prop}
\end{eqnarray}
where $E_i$ is the energy of the initial nuclei, $E_m$ is the energy of the
intermediate nuclei state $m$, $E_{ei}$, and $E_{\nu i}$ are the energies 
of electrons and neutrinos respectively. 
The minus signs between the two terms in the above expressions are 
due to the bosonic character of neutrinos;  in the case of  fermionic
neutrinos we would have 
plus signs~\cite{boehm}. For electrons we assume the normal Fermi 
statistics. 

In the case of   $0^+ \rightarrow 0^+$ transitions the combinations  $K_m$ 
and $L_m$ can be approximated by 
\be
K_m \approx \frac{E_{e2} - E_{e1} + E_{\nu2} - E_{\nu1}}{( E_m  - E_i   
+ E_0/2 )^2}, 
~~~~L_m \approx \frac{E_{e1} - E_{e2} + E_{\nu 2} - E_{\nu1}}{( E_m - 
E_i + E_0/2 )^2}, 
\label{KL}
\ee
whereas for the  fermionic neutrinos 
\be
K_m \approx L_m \approx  \frac{2}{E_m  - E_i + E_0/2 }.
\ee
Here $E_0/2 = E_e + E_{\nu}$ is the average energy of the leptonic pair. 
Appearance of the differences of the electron and neutrino 
energies 
in (\ref{KL}) leads to a suppression of the total probability. 
It also modifies the energy distributions of electrons.
The probabilities of the transitions $0^+ \rightarrow 2^+$
are proportional to the combinations $ (K_m - L_m)(K_n - L_n)$,
where
\be
(K_m - L_m) \approx  \frac{2(E_{e2} - E_{e1})}{(E_m  - E_i + 
E_0/2)^2}.
\ee
In the case of the fermionic neutrinos  
the combination has an additional factor $(E_{\nu2} - E_{\nu1})/(E_m  - 
E_i + E_0/2)$ and the suppression is stronger.

A large number of already detected events~\cite{klapdor} 
and especially future measurements~\cite{barabash} allow to make 
precision tests of the Pauli principle. The data  seems to exclude   
the 100\% violation of Fermi statistics for {\it electronic} 
neutrinos~\cite{prelim}. 
Notice however, that relation between the statistics of neutrinos and 
possible (small) violation of the Pauli principle is an open issue. \\

In what follows we will consider for simplicity an extreme case 
when neutrinos have purely bosonic statistics. 
In this case the equilibrium density of bosonic neutrinos in phase space 
would be equal to
\be
f_\nu = (2\pi)^3 n_0 \,\delta ({\bf p}) +
\left[ \exp \left( E-\mu\right) /T -1 \right]^{-1}, 
\label{f-nu}
\ee
where $\mu$ is the chemical potential of neutrinos. 
The first term describes a possible neutrino
condensate with $n_0$ being the number density of neutrinos in the 
condensate. The density (\ref{f-nu}) is somewhat larger than the 
fermionic one. The thermal part (second term) alone gives the 
energy density of bosonic neutrinos at $T\gg \mnu$ larger than
the fermionic one by the factor 8/7.

Expression (\ref{f-nu}) is the  equilibrium solution of the
kinetic equations since the collision integral with such a 
distribution function vanishes for non-zero $n_0$ if $\mu=\mnu$. 
For smaller values of $\mu$, $\mu < \mnu$,  the collision integral 
vanishes only if $n_0 = 0$. Therefore the condensate would be
non-vanishing only if the charge asymmetry of neutrinos is 
so large that the maximum chemical potential $\mu = \mnu$ is not
sufficient to provide such an asymmetry. 
We will check this explicitly in the next section for the situation 
during the BBN epoch. 

\section{Big bang nucleosynthesis}

The Big bang nucleosynthesis (BBN) is very  sensitive to the neutrino
statistics.

Let us first  show that  the distribution (\ref{f-nu}) is 
the  equilibrium solution of the 
kinetic equations.  For  the processes 
\be
e^- + p \lrar n + \nue,\,\,\, ~~~~ e^+ + n \lrar p + \bar{\nu}_e
\label{n-p-react}
\ee
relevant for the BBN nucleosynthesis, 
the collision integral is given by the integral over the phase space 
of the function
\be
F[f]=f_\nu f_n (1-f_e)(1-f_p) - f_e f_p (1-f_n) (1+ f_\nu), 
\label{ff}
\ee
if the invariance with respect to time reversal holds. 
Here $f_p$,  $f_n$ and  $f_e$ are the densities of 
protons, neutrons  and electrons correspondingly. 
We assume that they are described by 
the equilibrium Fermi-Dirac distributions~\footnote{Since  
the violation of the Pauli principle is communicated to other sectors 
of theory, one may expect some deviation from the Fermi-Dirac  
distribution for other fermions as well, but the effects are probably small.} 
\be
f_F = \left[ \exp \left( E-\mu\right) /T + 1 \right]^{-1},~~~~F = e,~p,~n. 
\label{f-f}
\ee
Notice that the neutrino distribution enters 
the second  them in (\ref{ff}) with plus sign.  

Inserting (\ref{f-nu}) and (\ref{f-f})  into  (\ref{ff}) 
we find that $F[f]$ is proportional to 
\begin{eqnarray}
F \propto   
\left[\exp\frac{1}{T}(E_\nu - \mu_{\nu}) - 1 \right]^{-1} \left[ 1 - 
\exp\frac{1}{T}(E_n + E_\nu - E_e - E_p  + \mu_e + \mu_p  - \mu_n 
-\mu_\nu)\right]\\
+ (2\pi)^3 n_0 \,\delta ({\bf p}) 
\left[1 -
\exp[\frac{1}{T}(E_n  - E_e - E_p  - \mu_n + \mu_e + \mu_p)] \right].   
\label{equil-cond}
\end{eqnarray}
The first (thermal) term  vanishes because of the conservations of energy, 
$E_e + E_p= E_n + E_\nu$, 
and the chemical potential: $ \mu_e + \mu_p =  \mu_n + \mu_\nu$.
(The condition for $\mu$ is true only in equilibrium.)
The second term in (\ref{equil-cond}) becomes zero if $\mu_{\nu} = 
m_{\nu}$. Indeed,  using the conservations of the energy and chemical 
potential we can rewrite the sum in the exponent as 
$E_n  - E_e - E_p  - \mu_n + \mu_e + \mu_p = - E_{\nu} + \mu_\nu 
= - m_{\nu}  + \mu_\nu$, where the last equality  is valid since  
${\bf p} = 0$.  
So,  for $\mu_{\nu} =  m_{\nu}$,   $F[f]$ vanishes and so does the 
collision integral, even for $n_0 \neq 0$. 
Thus,  the  usual equilibrium distributions remain true 
despite breakdown  of the spin-statistics relation.
If neutrinos obey mixed statistics their equilibrium distribution
can be written as a sum of the Bose and Fermi ones at least in the
case when a single neutrino participates in initial and/or final states 
of reactions.

If T-invariance is broken, then a more complicated expression appears in the
collision integral which still vanishes for the equilibrium distribution
because of the S-matrix unitarity~\cite{scl-bal}.

Since chemical potential of bosons cannot exceed their 
mass and neutrino mass is at most about eV, the chemical potential 
of neutrinos would not be essential in the BBN range of
temperatures, $T= (1 - 0.07) $ MeV. A possible large value of 
lepton charge asymmetry would be hidden in the condensate and
would not have a strong impact on chemical abundances of light
elements. If this is the case, the restrictive BBN 
bounds~\cite{LS,bbn-asym} on the magnitude of the cosmological lepton 
asymmetry would be inapplicable.

The equilibrium energy density of the  bosonic neutrinos at $T\gg \mnu$ is 
8/7 of the energy density of fermionic neutrinos and thus the change
of statistics would lead to an increase of the effective number of 
neutrino species at BBN by $\Delta N_\nu =3/7$ (for three neutrinos). 
On  the other hand, a  larger
magnitude of the neutrino distribution function and the fact that it
enters the kinetic equation (see (\ref{ff}))  as $(1+f_\nu)$ instead of  
$(1-f_\nu)$ makes 
the weak reactions of neutron-proton transformations (\ref{n-p-react}) 
faster and the $n/p$ freezing temperature becomes lower.      
The latter
dominates and as a result the effective number of massless species 
becomes smaller than 3.

To estimate the effect we proceed in the following simplified way.
The kinetic equation which governs the neutron-proton transformation
has the form (see, {\it  e.g.}, ref.~\cite{ad-nu-rev}):
\be
H x\frac{dr}{dx} = \frac{(1+3g^2_A) G_F^2}{2\pi^3} \left[A - (A+B)\,r 
\right], 
\label{dotr}
\ee
where $r$ is the ratio of neutron number density to the total baryon 
number density, $r \equiv n/(n+p)$, 
$x$ is the cosmological scale factor, 
$g_A = -1.267$ is the axial coupling constant, the
function $A(T)=B(T)\exp\left( -\Delta m/T\right)$, $\Delta m = 1.3$ MeV is
the neutron-proton mass difference. The function $B(T)$ is the 
collision integral containing the equilibrium distribution functions
of electrons and neutrinos.

If neutrinos have Bose-Einstein statistics, then the numerical value of 
$B(T)$ would differ from the standard fermionic one. The ratio of fermionic 
to bosonic $B$'s 
at the moment of neutron-proton freezing ($T\approx \Delta m/2$) is 
$B_f/B_b = 0.933$. This rise in the  reaction rate 
would result in a smaller $(n/p)$-freezing temperature. The same shift 
of the freezing temperature could be mimicked by a change
in the number of the effective
neutrino species which enters the expression for the Hubble parameter:
\be
H \sim \sqrt{g_*} = \sqrt{10.75 + 1.75 \Delta N_\nu}. 
\label{H-of-g}
\ee
One can check that the increase in the reaction rate due to the larger 
$B$ is equivalent to $\Delta N_\nu = -0.8$. 
Besides that one should take into 
account that a real change of the effective number of particle 
species changes not only the neutron-proton freezing temperature but also 
the moment when formation 
of the light elements begins. It takes place at about 0.07 MeV and with a 
smaller number of the particle species this temperature would be reached 
in a  longer  time and less neutrons would survive the decay. This effect 
brings about  20\% into the total impact on $\Delta N_{\nu}$ during BBN. 
Since the variation of $B$ does not change 
the time  of nucleus formation the total effect of this
change on $\Delta N_\nu$ should be $-0.6$ instead of 
$-0.8$ found above. 
Together with the positive contribution from three bosonic 
neutrinos into the total cosmological energy density, which is equivalent 
to
$\Delta N_\nu = 3/7$, we find that the bosonic neutrinos would make the 
effective number of neutrino species smaller by $\Delta N_\nu \approx -0.2$. 
The detailed calculations using properly modified 
BBN code give~\cite{hansen} $\Delta N_{\nu} = -0.94$ 
for the reaction rate change effect,   
and 
\be
\Delta N_\nu \approx -0.51  
\label{delta-N-nu}
\ee
for the total effect.

The neutrino condensate would not noticeably change the equilibrium 
value of $n/p$-ratio:
\be
(n/p)_{eq} = \exp \left[ -(\Delta m + \mu)/T \right] 
\label{n/p-eq}
\ee
because the chemical potential of neutrinos is 
negligible , $\mu = \mnu \ll T$.

For a positive neutrino asymmetry, 
{\it i.e.},  for $n_\nu > n_{\bar \nu}$ the condensate 
can contribute to the 
transition $\nu + n \rar p + e$, while for a negative asymmetry the 
contribution of the condensate to the $(n\lrar p)$-transformation is 
kinematically forbidden. These effects are automatically
included in the case of equilibrium, but when the $n/p$-ratio is out of 
equilibrium 
its evolution proceeds somewhat differently from the usual case,  
and additional processes with condensate should be taken into account.  
This  would make $\Delta N_\nu$ a few per cent more 
negative.

The negative value of $\Delta N_\nu$ 
found above seems to be in concordance with the data on the light 
element abundances. According to the recent analysis~\cite{cyburt}
$2.67 < N_\nu < 3.85$ at 68\% CL, while ref.~\cite{kneller} presents
the result $\Delta N_\nu = - 0.37 ^{+0.10}_{-0.11}$. Anyhow, taken 
at the face value,  the observed abundances of $^4$He and $^2$H  seem 
to be in contradiction with the standard BBN calculations, especially if
lower values of the observed abundances are taken. 
At the same time $\Delta N_\nu = -0.2$  opens some room for  
additional light degrees of freedom.

If neutrinos have mixed statistics, {\it e.g.}, the Pauli principle is 
broken for sterile neutrinos which mix  weakly with the standard ones,   
the effect of ``wrong'' statistics on $N_\nu$ would be further 
diminished.

\section{Bosonic neutrinos and dark matter}

If neutrinos obey the Bose-Einstein statistics, 
the Gunn-Tremaine~\cite{gt} lower bound on their mass would be 
inapplicable. This bound is based on the Fermi statistics which does not
permit to have too many neutrinos in a galaxy. Thus, to make all 
galactic dark matter, neutrinos (or any other fermions) must be
sufficiently heavy, $m_\nu \sim 100$ eV. However, this large  mass
contradicts the Gerstein-Zeldovich  upper bound~\cite{gz}. 
Bosonic neutrinos  can be arbitrary light, even as light
as axions, but still make all cosmological dark matter.

To play a noticeable role in the cosmological large scale structure (LSS) 
formation and, possibly, to make all the observed cold dark matter
the neutrino condensate should be sufficiently large. 
For $\mnu \approx 0.1$ eV 
the required energy density $\Omega_c = 0.25$ or 
$ \rho_c=1.25~ {\rm keV cm^{-3}}$ can be achieved if  
the number density of neutrinos in the condensate  
is  $1.25\cdot 10^4$ cm$^{-3}$.  
It is approximately 100 times larger than the number density of 
neutrinos plus antineutrinos for any single neutrino species from the 
cosmological thermal bath. 
To create such a large number of cold neutrinos the cosmological lepton
asymmetry must be much larger than 1. 

The large lepton  asymmetry
can be created in a version of the Affleck-Dine scenario~\cite{affl}.
Suppose  there is  a scalar field, $\chi$,  with non-zero lepton charge. 
This field can acquire a large vacuum expectation value
during inflation. The potential of $\chi$ self-interaction,  
$U(\chi)$, is supposed to 
break the leptonic charge conservation,  
and therefore $U(\chi)$ is not invariant 
with respect to phase transformation, $\chi\rar \chi \exp (i \beta)$. Usually
it is assumed to have the form 
\be
U(\chi) = m_1^2 |\chi|^2 + m^2_2 \chi^2 + m_2^{*2} \chi^{*2}
+ \lambda_1 |\chi|^4 + \lambda_2 \chi^4 + \lambda_2^{*} \chi^{*4}, 
\label{u-of-chi}
\ee
where $m_1$ and $\lambda_1$ are real but $m_2$ and $\lambda_2$ may be
complex so that the C-invariance is broken. 

If the mass parameters $m_{1,2}$
are small in comparison with the Hubble constant during inflation, $H_I$, 
then $\chi^2$ could acquire a large vacuum expectation value 
\be 
\langle \chi^2 \rangle  \sim \frac{H_I^4}{m^2_1}~~~ {\rm or} \sim  
\frac{H_I^2}{\lambda_1}. 
\label{vev}
\ee
Moreover, along possible flat
directions of the potential (valleys) $\chi^2$ would rise with time as
$H^3_I t$ (see the  discussion and the list of references, {\it  e.g.},  
in ~\cite{linde-book}).

The initial leptonic charge density of $\chi$ is equal to
\be
L_\chi = \dot \theta |\chi|^2,
\label{l-of-chi}
\ee
where $\theta$ is the phase of the field, $\chi = |\chi|\exp (i\theta)$.
The energy density of $\chi$ is about the kinetic energy,  
$\dot\theta^2 |\chi|^2$, and 
the potential energy is normally of the same order of 
magnitude. 

When inflation is over and the Hubble parameter becomes smaller than
the mass of $\chi$,  the field starts to relax down to zero and in the 
process of relaxation it transmits the accumulated leptonic charge into
leptonic charge of the decay products. 
We assume that $\chi$ decays into two neutrinos, 
$\chi \rightarrow \nu \nu$, (and not antineutrinos), {\it  i.e.},  
the leptonic charge of $\chi$ is two. 
In this case the lepton number density of the produced neutrinos would be 
\be
n_L = L_\chi. 
\label{eqeq}
\ee 
If the decay and the subsequent 
thermalization are fast, the energy density of the created plasma
becomes 
\be
\rho_T = \dot\theta^2 |\chi|^2 = (\pi^2 g_* /30) T^4, 
\label{rho-T}
\ee
where $g_*$ is the number of relativistic degrees of freedom in the plasma.

It follows from (\ref{l-of-chi}), 
(\ref{eqeq}) and (\ref{rho-T}) that the  charge  
asymmetry of the plasma could be
\be
\beta_L = n_L/T^3 \sim \sqrt {\chi /\dot \theta}.  
\label{nL-T3}
\ee
For the  initial values
$\dot\theta \sim H_I$ and $\chi \sim H^2_I/m$,  the asymmetry equals 
\be 
\beta_L \sim \frac{H}{m}, 
\ee
and it does not look unreasonable, {\it e.g.}, 
that $\beta_L>100$.  

After $\chi$ decayed into neutrino pairs, the plasma thermalized through
the reactions $\nu\nu \rar \nu\nu l\bar l$, {\it etc.}.  Complete 
thermalization would produce the equilibrium distribution of neutrinos 
given by eq. (\ref{f-nu}). 
Because of a low rate of reactions with
zero momentum neutrinos the complete equilibrium with condensation at 
${\bf p}_\nu =0$ is never reached, still, according to the estimates presented
below, neutrinos accumulate sufficiently close to ${\bf p}=0$ and 
the condensate practically forms.

The kinetic equilibrium could be achieved 
through elastic scattering reactions $\nu+l \rar \nu +l$. If the initial
temperature was above the electroweak scale and therefore the intermediate
$W$ and $Z$ bosons could be considered massless, the low energy band of
the spectrum of neutrinos would be filled in with the rate
\be 
\Gamma \sim g^4 E_\nu \sim \alpha^2 E_\nu,
\label{Gamma}
\ee
where $g$ is the electroweak coupling constant, $E_{\nu}$ is the 
neutrino energy and $\alpha = 1/137$.
At temperatures below the masses of $W$ and $Z$ bosons, the rate becomes 
much lower: 
\be 
\Gamma \sim G_F^2 T^4 E_\nu,
\label{Gammab}
\ee
where $G_F$ is the Fermi coupling constant.

Requiring $\Gamma\geq H$, we conclude that neutrinos ``condense'' down to 
energies $E_\nu/T \sim 10 T/m_{Pl}$ if the process occurs before  the EW 
phase transition, i.e. if $\chi$ decays at temperatures $T > m_{W,Z}$.  
When $T$ drops below $m_{W,Z}$
the cooling would become 
less efficient. Effective energies to which neutrinos could cool down  
in the condensation process increases up to $m_W^4 /(T^2 m_{Pl})$.
Though complete condensation with ${\bf p} =0$ could 
probably never be achieved,
an accumulation of neutrinos at low energy part of the spectrum seems to
be efficient enough to create cosmological cold dark matter of neutrinos  
\footnote{In the case that sterile neutrinos condense, the necessary for DM 
magnitude of the condensate should be scaled by the ratio of 
sterile/active masses.}.\\

In contrast to the case of large charge asymmetry carried by fermions 
(when the spectrum is populated by energetic particles),   in the bosonic
case the charge asymmetry is stored in the   condensate, while
chemical potential $\mu=\mnu$ is very small and
the energy density is dominated by relativistic particles with
practically vanishing charge asymmetry. It remains 
true till temperature dropped down below $\mnu$. After that the
energy density would be dominated by non-relativistic condensate with
large leptonic charge. 

The energy density in the neutrino condensate equals 
\be
\rho_{cond} = m_{\nu} n_0 .
\label{edens}
\ee
Requiring that at the BBN epoch $\rho_{cond} < 0.1 \rho_{\nu}^T$, 
where $\rho_{\nu}^T$ is the energy density in the thermal bath,  
we find the bound
\be
n_0 < 0.3~ n_{\nu}^T  \frac{T_{BBN}}{m_{\nu}} \sim 10^{6} n_{\nu}^T . 
\label{bound}
\ee
Here $n_{\nu}^T$ is the neutrino number density in the 
thermal bath.  

If the equilibrium concentration (\ref{f-nu}) is not achieved and 
the condensate is not formed completely, so that ${\bf p} \neq 0$, 
the BBN gives the following bound: 
\be
|{\bf p}| < 0.3~ T_{BBN} \frac{n_{\nu}^T}{n_0}. 
\ee
For $n_{\nu}^T / n_0 \sim 10^{-2}$ required to explain the 
dark matter we find $|{\bf p}| < 3\cdot 10^3$ eV. \\

The large scale structure formation with cold dark matter composed of the 
neutrino condensate would well fit the observed picture. The analysis
of structure formation with the Bose condensate of normal bosons was done
in ref.~\cite{madsen} and it can be extended to the case of neutrino
without significant modifications.\\

One can be less ambitious and admit  two coexisting forms of the 
cold dark matter: {\it e.g.},  the usually accepted lightest 
supersymmetric  particle (LSP) or  axions, and the Bose-condensed
neutrinos. It would make the scenario considered here
less vulnerable but simultaneously less predictive.

Notice that because of  smallness of the neutrino chemical potential
a large lepton asymmetry in primordial plasma cannot be transformed 
into the baryon one by electroweak processes. Indeed, the chemical
potential of neutrinos is tiny and in equilibrium the same must
be true for chemical potentials of quarks. Hence the baryon asymmetry
generated by sphalerons in equilibrium could be at most of the order  
$\mnu/T \sim 10^{-12}$.

\section{Astrophysical consequences}

The neutrino statistics plays the key role in the environments where 
neutrinos form dense degenerate gases.

Direct test of the ``bosonic'' nature of neutrinos can be probed by 
precise measurements  of the neutrino energy  spectrum from supernova. 
Instead of the Fermi-Dirac spectrum with pinching the distribution would be
the Bose-Einstein one with some pinching effect too. So, generically,  
the spectrum  of bosonic neutrinos should be narrower. 
To establish the difference one  needs to measure 
the spectrum both in the low, 
$E < 3T$,  and in the high, $E > 3T$ energy parts.  Also pinching effect 
should be quantified rather precisely. 

Violation of the Pauli principle can  influence  
dynamics of the SN collapse. According to the usual scenario 
in the initial stages (formation of the hot proto-neutron star)  
the neutronization  
leads to production of high concentration of the electron neutrinos 
which are trapped in the core. The chemical potential of these neutrinos 
(due to the Pauli principle) can reach 70 - 100 MeV. These neutrinos heat 
the 
medium and diffuse from  the core. Violation of the Pauli principle allows 
for  the neutronization  neutrinos to be produced with lower energies.  
These neutrinos escape easier the star leading to faster cooling 
and lower central temperatures. 
Also the evolution of the lepton number would change. 

The presence of the neutrino condensate with large lepton 
number in the Universe  may have a number of observable consequences. 

High neutrino density in the condensate
(especially if an additional clustering occurs)
enhances rate of the  $Z^0$-bursts produced by  annihilation of the 
ultra high energy (UHE)  cosmic 
neutrinos on the relic neutrinos~\cite{tw,far}. 
This in turn, enhances  production of the UHE
cosmic rays, and  may help to explain the cosmic ray evens above  the
GZK cut-off. 

The asymmetric neutrino condensate may produce strong refraction  of   
the high energy neutrinos from remote sources (active galactic nuclei, gamma 
ray bursters).  Apart from lensing, one may expect a substantial 
impact on neutrino oscillations~\cite{LS}. 

Since the density of dark matter in galaxies is about 6 orders of 
magnitude larger than their average cosmological energy density, 
a condensation of cold neutrinos around the Earth might have an 
effect on the end point of the beta decay spectra,  in particular, in the 
tritium decay experiments on search for neutrino mass.\\

\section{Conclusions}

If the Pauli principle is violated in Nature, neutrinos may be 
the first messengers of that. Indeed,  neutrino properties
may be  related to physics at high energy  
scales and only neutrinos (due to their neutrality) can mix with
particles of the hidden sector (singlets of the standard model 
symmetry group).\\

\noindent
We have considered the cosmological and astrophysical consequences
of the Pauli principle violation assuming that
neutrinos obey the Bose-Einstein statistics. 
Such neutrinos would have the Bose-Einstein energy distributions
and might form a condensate.\\
     
\noindent
We have found that in the case of bosonic neutrinos,
the effective number of the neutrino species  during the BBN epoch is 
reduced by about $\Delta N_{\nu} = - 0.5$, 
thus slightly facilitating an existence of new degrees 
of freedom, {\it e.g.},  sterile neutrinos.\\  

\noindent
If in the Early Universe,  a large lepton asymmetry was created, bosonic
neutrinos should condense in the process of the cosmological cooling.
Because of that the BBN bound on the neutrino density can be avoided, and so 
a large neutrino concentration and energy density at a later epoch become
possible. In  particular, it allows to
explain all or a part of the cosmological dark matter. 
It opens a possibility of large lepton asymmetry of the Universe. \\

\noindent
Bosonic neutrinos would influence dynamics of the stellar collapse
and lead to modification of the energy spectra of SN neutrinos.\\

\noindent
A large concentration of relic background neutrinos in the condensate
at the present epoch would enhance the rate of the $Z^0-$bursts.
It can produce strong refraction effect on neutrinos from remote 
sources, modifying the oscillation pattern. \\ 

\noindent
A study of the double beta decay, in particular, 
of the energy spectrum and  angular distribution of electrons from the 
decay can provide sensitive test of the Pauli principle violation and 
statistics of neutrinos.\\

\bigskip

\noindent
{\bf Note added:} After submission of the first version of our paper, 
we became aware  of the paper \cite{gri} in which a possibility of bosonic 
statistics for neutrinos has been considered and its effects on the BBN have been 
studied. According to \cite{gri} the change of neutrino 
statistics is equivalent to the decrease of number of 
the effective neutrino species $\Delta N_{\nu} = -0.74$. 
This conclusion agrees qualitatively with our results of sec.~3,  
though quantitatively we find smaller decrease.

\section{Acknowledgments}

We are grateful to S.I. Blinnikov and L.B. Okun for helpful discussions.
We thank S. Pastor who informed us about ref.~\cite{gri}.

\end{document}